\documentclass[11pt, letterpaper]{article}

\usepackage[margin=2.54cm]{geometry}
\usepackage{lineno}
\usepackage{amsmath,amssymb}
\usepackage{graphicx}
\usepackage[round,comma,authoryear]{natbib}
\bibpunct[]{(}{)}{,}{a}{}{,}
\usepackage{float}
\usepackage{setspace}
\usepackage{comment}
\usepackage{booktabs}
\usepackage{lscape}
\usepackage{titlesec}
\usepackage{rotating}
\usepackage{nidanfloat}
\usepackage{color}

\newcommand {\thlfh} {0.5 - (\theta_{it} - \lfloor \theta_{it} \rfloor)}
\newcommand {\thllh} {\lceil \theta_{it} \rceil - \theta_{it}}

\titleformat*{\section}{\normalsize\bfseries}
\titleformat*{\subsection}{\normalsize\bfseries}

\title{
A self-excited threshold autoregressive state-space model for menstrual cycles: forecasting menstruation and identifying ovarian phases based on basal body temperature
}

\begin{small}
\author{
  {\small Ai Kawamori,$^{1}$ Keiichi Fukaya,$^{1*}$ 
    Masumi Kitazawa$^{2}$ and Makio Ishiguro$^{1}$} \\\\
  {\footnotesize \it $^{1}$The Institute of Statistical Mathematics, 10-3 Midoricho, Tachikawa, Tokyo 190-8562 Japan} \\
  {\footnotesize \it $^{2}$QOL Corporation, 2929, Netsu, Tomi, Nagano 389-0506, Japan} \\
  {\footnotesize $^{*}$kfukaya@ism.ac.jp}
}
\end{small}

\begin{document}

\date{ }

\begin{spacing}{1.2}
\maketitle

\begin{abstract}
  The menstrual cycle is composed of the follicular phase and subsequent luteal phase
  based on events occurring in the ovary.
  Basal body temperature (BBT) reflects this biphasic aspect of menstrual cycle and
  tends to be relatively low during the follicular phase. 
  In the present study, we proposed a state-space model that explicitly incorporates
  the biphasic nature of the menstrual cycle, in which the probability density distributions
  for the advancement of the menstrual phase and that for BBT switch depend on a
  latent state variable.
  Our model derives the predictive distribution of the day of the next menstruation onset 
  that is adaptively adjusted by accommodating new observations of BBT sequentially.
  It also enables us to obtain conditional probabilities of the woman being in the early or
  late stages of the cycle, which can be used to identify the duration of follicular and
  luteal phases, as well as to estimate the day of ovulation.
  By applying the model to real BBT and menstruation data, we show that
  the proposed model can properly capture the biphasic characteristics of menstrual cycles,
  providing a good prediction of the menstruation onset in a wide range of age groups.
  An application to a large data set containing 25,622 cycles provided by 3,533 woman subjects
  further highlighted the between-age differences in the population characteristics of
  menstrual cycles, suggesting wide applicability of the proposed model. \\

\end{abstract}

\section{Introduction}

During the reproductive age, women experience recurring physiological changes
known as menstrual cycles.
The cycle starts on the first day of menstruation, followed by a pre-ovulatory period
referred to as the follicular phase.
After ovulation, the cycle enters a post-ovulatory period referred to as the luteal phase,
lasting until the day before the next menstruation onset.
Although menstrual cycles generally last 28 days, the length of the menstrual cycle
exhibits significant variation, both within and among individuals \citep{Harlow1991}.
Variation in menstrual cycle length is mainly attributed
to the follicular phase, as the follicular phase shows greater variation in length
than the luteal phase \citep{Fehring2006}.
Thus, determining the time of ovulation can be difficult.

Basal body temperature (BBT) also reflects this biphasic aspect of the menstrual cycle;
BBT tends to be relatively low during the follicular phase, 
increasing by $0.3$ to $0.5\ {}^\circ\mathrm{C}$ after the cycle enters the luteal phase
\citep{Barron2005, Scarpa2009}.
Since a shift in BBT may be indicative of ovulation, daily BBT records could be
used to estimate the day of ovulation and associated fertile interval.
However, the estimation of ovulation based on BBT may be error-prone \citep{Barron2005,Dunson2000}.

Considerable effort has been made to develop menstrual cycle-related
statistical models.
Barrett-Marshall-Schwartz models are a class of statistical models for human fecundability,
in which the occurrence of pregnancy is explained by the intercourse pattern and 
day-specific probability of conception within the fertile interval
\citep{Barrett1969, Schwartz1980}.
The fertile interval within the menstrual cycle refers to the time period where
the day-specific probability of conception is not negligible;
thus, intercourse can result in pregnancy.
It is estimated to last about 6 days,
starting $\sim5$ days prior to ovulation and ending on the day of ovulation
\citep{Dunson1999,Dunson2002}.
The fertile interval can be identified based on various biological markers for ovulation,
such as BBT, urinary luteinizing hormone level, and cervical mucus.
However, none of these markers identify ovulation perfectly;
therefore, error in the identification of the day of ovulation is
considered an important issue in studies of human fecundability
\citep{Dunson2000, Dunson2001}.
Time-to-pregnancy models are 
another class of statistical models for human fecundability, 
which explain the number of menstrual cycles required to achieve a
clinical pregnancy.
Additional statistical models for human fecundability have been reviewed in
\cite{Weinberg2000,Ecochard2006,Zhou2006,Scarpa2010}, and \cite{Sundaram2010}.

Another line of research involves the development of statistical models 
explaining the variability in the menstrual cycle length (MCL).
This includes the following types of models:
mixture distribution models explaining the long right tail in the distribution of the MCL
\citep{Harlow1991,Guo2006,McLain2012};
longitudinal models, accounting for the within subject-correlation in the MCLs
\citep{Harlow1991,Lin1997,McLain2012};
and a change point model identifying the menopausal shift in the moments of the MCL distribution
\citep{Huang2014}.
Furthermore, \cite{Bortot2010} proposed a state-space modeling framework
providing a predictive distribution of the MCL that is conditional on
the past time series of the MCL.
Moreover, recent studies have considered the joint modeling of the MCL and fecundability
\citep{McLain2012,Lum2016,Liu2017}.

Although BBT is an easily observed quantity relevant to the menstrual cycle,
the development of statistical models explaining
periodic BBT fluctuations has received little attention.
\cite{Scarpa2009} applied Bayesian functional data analysis
to BBT time series data
that characterized BBT fluctuations in normal cycles parametrically 
and identified abnormal BBT trajectories nonparametrically.
\cite{Fukaya2017} recently proposed a state-space model
involving the menstrual phase as a latent state variable
explaining the BBT time series.
Applying a sequential Bayesian filtering algorithm enabled the authors to obtain
filtering distribution of the menstrual phase, providing a predictive distribution
for the onset of the next menstruation sequentially.

In the model proposed by \cite{Fukaya2017}, the biphasic nature of the
menstrual cycle was not accounted for in an explicit manner.
Specifically, the model used a trigonometric series to explain
the periodic BBT fluctuations, and
the biphasic pattern may thus appear as a result of model fitting.
In addition, the model assumed a single probability density distribution for the
advancement of the menstrual phase, and did not account for differences in
the distribution of the length of the follicular and luteal phases.
However, there is 
another possible model formulation that is biologically more natural and interpretable,
based on previous knowledge of the menstrual cycle.
This model involves 
dividing a cycle into two distinct stages (i.e., first and second stages,
which are expected to correspond to the follicular and luteal phases, respectively),
in which each stage is characterized by specific statistical distributions for BBT and
the advancement of the menstrual phase.

In the present study, we propose such an ``explicit biphasic menstrual cycle model'',
as an extension of the model proposed by \cite{Fukaya2017},
which we refer to as an ``implicit biphasic menstrual cycle model''.
In our explicit model, the probability density distributions of
the advancement of the menstrual phase
and the BBT switch, depending on a latent state variable.
Our model can therefore be seen as a self-excited threshold autoregressive
state-space model \citep{Ives2012}.
Most of the statistical inferences applied to the implicit model described in the
previous paper \citep{Fukaya2017} can be similarly applied to the current explicit model.
Thus, the conditional distribution of the latent menstrual phase variable
can be obtained using a sequential Bayesian filtering algorithm,
which in turn can be used to yield a predictive distribution of the
day of menstruation onset.
Furthermore, as we describe below, the conditional distribution of the
latent menstrual phase variable
naturally provides the probability of a subject being in the first or second stage,
which may potentially be used to estimate ovulation, provided the model adequately captures
the characteristics of the follicular and luteal phases.
By applying to a large data set of menstrual cycles, we illustrate the
wide applicability of the proposed model.
We show that the proposed model can properly capture the biphasic characteristics of
menstrual cycles, shedding light on the between-age differences in the
population characteristics of menstrual cycles.

The remainder of this paper is organized as follows.
In Section \ref{sect:model}, we detail the proposed method,
formulating the model and describing statistical inferences 
involving latent state variables and parameters.
We also explain how the predictive distribution for the next menstruation onset
and the probability that a subject is in the first or second stage
can be obtained, based on the conditional distribution of the menstrual phase variable.
Section \ref{sect:appl} presents an application of the proposed model to a
real menstrual cycle data set
collected from a large number of women.
For eight age groups, ranging from the late teens to the early 50s, 
we report the maximum likelihood estimates of the model parameters, and
examine the accuracy of the prediction of menstruation onset.
We also provide the joint and marginal distributions of the lengths of the
first and second stages,
which was judged based on the smoothed distribution of the menstrual phase.
A concluding discussion is provided in Section \ref{sect:disc}.

\section{Model description and inferences}
\label{sect:model}

\subsection{State-space model of the menstrual cycle}
\label{sect:ssm}

Suppose for $i = 1, \dots, I$ female subjects, 
a record of BBT measurement, $y_{it}$, and an indicator of the onset of menstruation, $z_{it}$,
was obtained for days $t=1,\dots,T_i$.
By $z_{it} = 1$, we denote that menstruation for subject $i$ started on day $t$,
whereas $z_{it} = 0$ indicates that day $t$ was not the first day of menstruation for the
subject $i$.
We denote the BBT time series and menstruation data obtained from the subject $i$
up to time $t$ as
$Y_{it} = (y_{i1}, \dots, y_{it})$ and $Z_{it} = (z_{i1}, \dots, z_{it})$, respectively.
We assumed that the time series for each subject was independent from the 
time series of other subjects.

We considered the phase of the menstrual cycle,
$\theta_{it} \in \mathbb{R}$,
to be a latent state variable.
In the following, we assumed that menstrual cycles are 
periodic in terms of $\theta_{it}$ with a period of 1.
We divided each cycle into two distinct stages,
the first stage
($0 \leq \theta_{it} - \lfloor \theta_{it} \rfloor < 0.5$)
and the second stage
($0.5 \leq \theta_{it} - \lfloor \theta_{it} \rfloor < 1$),
where $\lfloor x \rfloor$ is the floor function returning
the largest previous integer for $x$.
We expected the former to represent the follicular phase and 
the latter to represent the luteal phase.
We defined a set of real numbers corresponding to the latent menstrual phase being 
in the first and second stages of the cycle as
$\Theta_1 = \{ \theta \subset \mathbb{R} \mid 0 \leq \theta - \lfloor \theta \rfloor < 0.5 \}$
and $\Theta_2 = \{ \theta \subset \mathbb{R} \mid 0.5 \leq \theta - \lfloor \theta \rfloor < 1 \}$,
respectively.

We let $\epsilon_{it}$ reflect the daily advance of the phase
for subject $i$ between days $t-1$ and $t$,
and assumed that it was a positive random variable independently 
following a gamma distribution
with varying parameters.
Thus, the system model can be described as:
%
\begin{align}
  & \theta_{it} = \theta_{i,t-1} + \epsilon_{it} \\
  & \epsilon_{it} \sim \textrm{Gamma}\left\{\alpha\left(\theta_{i,t-1}\right), \beta\left(\theta_{i,t-1}\right) \right\} \\
  & \left\{\alpha\left(\theta_{i,t-1}\right), \beta\left(\theta_{i,t-1}\right) \right\} = \begin{cases} 
      (\alpha_{i1}, \beta_{i1}) \hspace{1em} \textrm{when} \hspace{1em}
      \theta_{i,t-1} \in \Theta_1 \\
      (\alpha_{i2}, \beta_{i2})  \hspace{1em} \textrm{when} \hspace{1em}
      \theta_{i,t-1} \in \Theta_2. \\
    \end{cases}
\end{align}
We assumed that the system model parameters
switched between two stages,
enabling a description of the difference
in the variability of the length of these stages.
Under this assumption, the conditional distribution of $\theta_{it}$, given $\theta_{i,t-1}$,
is a gamma distribution with a probability density function:
%
\begin{align}
  p(\theta_{it} \mid \theta_{i,t-1}) &= \textrm{Gamma}\left\{\alpha\left(\theta_{i,t-1}\right), \beta\left(\theta_{i,t-1}\right) \right\} \notag \\
  & = \frac{\beta\left(\theta_{i,t-1}\right)^{\alpha\left(\theta_{i,t-1}\right)}}{\Gamma\left\{\alpha\left(\theta_{i,t-1}\right)\right\}}
    (\theta_{it} - \theta_{i,t-1})^{\alpha\left(\theta_{i,t-1}\right) - 1} 
    \exp \left\{-\beta\left(\theta_{i,t-1}\right) (\theta_{it} - \theta_{i,t-1}) \right\}.
\end{align}

We assumed that the distribution of the observed BBT, $y_{it}$,
was conditional on the menstrual phase $\theta_{it}$.
Assuming a Gaussian observation error, the observation model for the BBT
can be expressed as:
%
\begin{align}
  &y_{it} = \mu\left(\theta_{it}\right) + e_{it} \\
  &e_{it} \sim \textrm{Normal}\left\{0, \sigma^2\left(\theta_{it}\right)\right\} \\
  &\left\{\mu\left(\theta_{it}\right), \sigma^2\left(\theta_{it}\right)\right\} = \begin{cases} 
      (\mu_{i1}, \sigma^2_{i1}) \hspace{1em} \textrm{when} \hspace{1em}
      \theta_{it} \in \Theta_1 \\
      (\mu_{i2}, \sigma^2_{i2})  \hspace{1em} \textrm{when} \hspace{1em}
      \theta_{it} \in \Theta_2. \\
    \end{cases}
\end{align}
Again, observation model parameters
$\left\{\mu\left(\theta_{it}\right), \sigma^2\left(\theta_{it}\right)\right\}$
switched depending on
the underlying stage within the cycle, in order to describe
a biphasic pattern in the BBT.
Conditional on $\theta_{it}$, $y_{it}$ then follows a normal distribution with a
probability density function:
%
\begin{align}
  p(y_{it} \mid \theta_{it}) &= \textrm{Normal}\left\{ \mu\left(\theta_{it}\right), \sigma^2\left(\theta_{it}\right) \right\} \notag \\
  & = \frac{1}{\sqrt{2\pi\sigma^2\left(\theta_{it}\right)}}
  \exp \left[ -\frac{\left\{ y_{it} - \mu\left(\theta_{it}\right)\right\}^2}{2\sigma^2\left(\theta_{it}\right)} \right].
\end{align}

For the menstruation onset, we assumed that 
menstruation starts when $\theta_{it}$ crosses the smallest following integer.
This can be represented as follows:
%
\begin{align}
  z_{it} = \begin{cases}
  0 \hspace{1em} \textrm{when} \hspace{1em}
    \lfloor \theta_{it} \rfloor = \lfloor \theta_{i,t-1} \rfloor \\
  1 \hspace{1em} \textrm{when} \hspace{1em}
    \lfloor \theta_{it} \rfloor > \lfloor \theta_{i,t-1} \rfloor. 
  \end{cases}
\end{align}
In rewriting this deterministic allocation in a probabilistic manner, 
$z_{it}$ follows a Bernoulli distribution conditional on $(\theta_{it}, \theta_{i,t-1})$:
%
\begin{align}
  p(z_{it} \mid \theta_{it}, \theta_{i,t-1}) 
  &= (1 - z_{it})\left\{I(\lfloor \theta_{it} \rfloor = \lfloor \theta_{i,t-1} \rfloor)\right\}
  + z_{it}\left\{I(\lfloor \theta_{it} \rfloor > \lfloor \theta_{i,t-1} \rfloor)\right\},
\end{align}
%

\noindent{where} $I(x)$ is the indicator function that returns 1 when $x$ is 
true or 0 otherwise.

The model involves, under no restriction, a total of $8 \times I$ parameters;
that is, the model has an independent set of parameters
$(\alpha_{i1}, \alpha_{i2},\beta_{i1}, \beta_{i2},\mu_{i1}, \mu_{i2},\sigma_{i1}, \sigma_{i2})$
for each subject $i$.
However, in most cases, it is likely that data are not sufficiently abundant
to estimate parameters separately for each subject.
Restricted versions of the model can be considered
by assuming that certain parameters are equal among subjects.
For example, we can assume that a set of parameters
$(\alpha_{i1}, \alpha_{i2},\beta_{i1}, \beta_{i2},\mu_{i1}, \mu_{i2},\sigma_{i1}, \sigma_{i2})$
is equal for all $I$ subjects, in which case there would be only 8 parameters to be estimated.
This approach pools information across subjects and allows the
estimation of parameters, even when the time series is not of sufficient length for each subject.
Between-subject variation can be accounted for 
using covariate information, such as the age of the subject, when available.
For example, between-subject differences in $\alpha_{i1} (> 0)$ can be
modeled as: $\log \alpha_{i1} = \gamma_0 + \sum_j \gamma_j x_{ij}$,
where $x_{ij}$ is the $j$th covariate for subject $i$.
In this case, the $\gamma$s are the parameters to be estimated.

Let $\boldsymbol{\xi}$ be a vector of the parameters of the model.
Given data $(Y_{iT_i}, Z_{iT_i})$ and a distribution specified for initial states
$p(\theta_{i1}, \theta_{i0})$ for each subject $i$,
the parameters in $\boldsymbol{\xi}$ can be estimated using the maximum likelihood method.
The log-likelihood for subject $i$ can be expressed as
%
\begin{align}
  &l_i(\boldsymbol{\xi}; Y_{iT_i}, Z_{iT_i}) 
  =\log p(y_{i1},z_{i1} \mid \boldsymbol{\xi})+
  \sum_{t=2}^{T_i}\log p(y_{it},z_{it} \mid Y_{i,t-1}, Z_{i,t-1}, \boldsymbol{\xi}),
\end{align}
\noindent{where}
%
\begin{align}
  &\log p(y_{i1},z_{i1} \mid \boldsymbol{\xi}) 
  = \log \int\int p(y_{i1} \mid \theta_{i1})p(z_{i1} \mid \theta_{i1}, \theta_{i0})p(\theta_{i1}, \theta_{i0}) d\theta_{i1} d\theta_{i0}.  \label{eqn:loglik1}
\end{align}
\noindent For $t = 2, \dots, T_i$,
%
\begin{align}
  &\log p(y_{it},z_{it} \mid Y_{i,t-1}, Z_{i,t-1}, \boldsymbol{\xi}) \notag \\
  &= \log \int\int p(y_{it} \mid \theta_{it})p(z_{it} \mid \theta_{it}, \theta_{i,t-1}) 
  p(\theta_{it}, \theta_{i,t-1} \mid Y_{i,t-1}, Z_{i,t-1}) d\theta_{it} d\theta_{i,t-1},  \label{eqn:loglikt}
\end{align}
which can be sequentially obtained using the Bayesian filtering technique described below.
Note that, for each subject $i$, $p(y_{it} \mid \theta_{it}) (t\geq1)$ and 
$p(\theta_{it}, \theta_{i,t-1} \mid Y_{i,t-1}, Z_{i,t-1}) (t\geq2)$ depend on $\boldsymbol{\xi}$;
however, this dependence is not explicitly described here for notational simplicity.
Since the subject time series is assumed to be independent,
the joint log-likelihood is the sum of the $I$ subject log-likelihoods:
%
\begin{align}
  l(\boldsymbol{\xi}; Y_{1T_1},\dots,Y_{IT_I}, Z_{1T_1},\dots,Z_{IT_I})=
  \sum_{i=1}^I l_i(\boldsymbol{\xi}; Y_{iT_i}, Z_{iT_i}).
\end{align}

\subsection{State estimation and calculation of log-likelihood by using sequential Bayesian filtering}
\label{sect:ngf}

The joint distribution of the phase of subject $i$ at successive time points $t$ and $t-1$,
conditional on the observations obtained up to time $u$,
$p(\theta_{it}, \theta_{i,t-1} \mid Y_{iu}, Z_{iu})$,
is referred to as a predictive distribution when $t > u$, 
as a filtering distribution when $t = u$, and as a smoothed distribution when $t < u$.
Given a state-space model, its parameters, and data,
these conditional distributions can be obtained by using 
recursive formulae for the state estimation problem,
which are referred to as the Bayesian filtering and smoothing equations.
Details regarding the state estimation of state-space models have been previously described
(e.g., \citealt{Kitagawa2010,Sarkka2013}).

Although these conditional probability distributions are often analytically intractable
for non-linear, non-Gaussian state-space models,
they can be obtained approximately for the self-excited threshold autoregressive
state-space model described above.
To this end, we applied Kitagawa's non-Gaussian filtering procedure \citep{Kitagawa1987},
which provides a numerical approximation of the joint conditional probability density
$p(\theta_{it}, \theta_{i,t-1} \mid Y_{iu}, Z_{iu})$.
The numerical procedure applied to the implicit biphasic menstrual cycle model
described by \cite{Fukaya2017}
can be used to our explicit biphasic menstrual cycle model in the same manner.
Since the subject time series are assumed to be independent,
conditional probability distributions can be obtained separately for each subject $i$.
Once the numerical approximation of the joint conditional probability density
$p(\theta_{it}, \theta_{i,t-1} \mid Y_{iu}, Z_{iu})$ is obtained,
the marginal probability densities
(e.g., $p(\theta_{it} \mid Y_{iu}, Z_{iu}) = \int p(\theta_{it}, \theta_{i,t-1} \mid Y_{iu}, Z_{iu}) d\theta_{i,t-1}$)
can also be obtained straightforwardly;
these are used to obtain the predictive distribution for the day of menstruation onset
and the conditional probabilities for the ovarian cycle phases as described below.
The log-likelihood for data at a particular time 
(Equations \ref{eqn:loglik1} and \ref{eqn:loglikt})
can be calculated for each subject as a by-product of obtaining the filtering distribution.
Missing observations are allowed in the sequential Bayesian filtering procedure.
More details can be found in \cite{Fukaya2017}.

\subsection{Sequential Bayesian prediction for the day of menstruation onset}
\label{sect:pred}

\cite{Fukaya2017} reported that
the filtering distribution of the menstrual phase in their state-space model
could be used to obtain
the predictive distribution for the next menstruation onset
(i.e., a predictive distribution for the length of the current cycle),
which was conditional on the accumulated data available at the time point of the prediction.
Since filtering distributions of the menstrual phase can be obtained using the
Bayesian filtering procedure,
the prediction of the day of onset for the next menstruation
can be adaptively adjusted by accommodating new observations sequentially.
This sequential predictive framework can be applied to
the current explicit model in the same manner; however,
the detailed calculations are more complicated due to 
the biphasic nature of the system model.

We denote, for $k = 1, 2, \dots$, the probability that the next menstruation of subject $i$
occurs on the $t + k$th day, conditional on the data available for her at the $t$th day
as $h(k \mid Y_{it}, Z_{it})$.
Given the marginal filtering distribution for the phase state
$p(\theta_{it} \mid Y_{it}, Z_{it})$, the conditional probability is as follows:
%
\begin{equation}
  h(k \mid Y_{it}, Z_{it}) = \int f(k \mid \theta_{it}) p(\theta_{it} \mid Y_{it}, Z_{it}) d\theta_{it},
\end{equation}
where $f(k \mid \theta_{it})$ is the conditional probability function for the
day of menstruation onset \citep{Fukaya2017}.
We provide details for calculation of $f(k \mid \theta_{it})$ 
under the proposed explicit biphasic menstrual cycle model
in Appendix \ref{sect:appa}.
Briefly, although $f(k \mid \theta_{it})$ can be obtained for $\theta_{it} \in \Theta_2$
straightforwardly, 
it is more involved for $\theta_{it} \in \Theta_1$.
This is because the probability that 
the first stage of the cycle lasts until the $t + j -1$th day
and the next menstruation occurs on the $t + k$th day,
denoted as $\phi(j,k \mid \theta_{it})$ ($j = 1, \dots, k$),
is needed to obtain $f(k \mid \theta_{it})$
such that $f(k \mid \theta_{it}) = \sum_{j=1}^k \phi(j,k \mid \theta_{it})$.
Furthermore, the calculation of $\phi(j,k \mid \theta_{it})$ requires
convolutions of probability distributions, rendering it being approximated numerically.

We can choose the $k$ giving the highest probability, $\max h(k \mid Y_{it}, Z_{it})$,
as a point prediction for the day of menstruation onset of subject $i$.

\subsection{Identification of stages of the cycle}
\label{sect:phase}

For some $t$ and $u$, the probability that the menstrual phase of subject $i$ at time $t$
is in the first stage of the cycle, conditional on data obtained by time $u$, can be given as
%
\begin{equation}
  \textrm{Pr}(\theta_{it} \in \Theta_1 \mid Y_{iu}, Z_{iu})
  = \int_{\Theta_1} p(\theta_{it} \mid Y_{iu}, Z_{iu}) d\theta_{it},
\end{equation}
which we expect to represent the probability of the subject $i$ being in the follicular phase
at time $t$.
By contrast, the conditional probability that the menstrual phase is in the
second stage of the cycle can be given as
%
\begin{align}
  \textrm{Pr}(\theta_{it} \in \Theta_2 \mid Y_{iu}, Z_{iu})
  &= \int_{\Theta_2} p(\theta_{it} \mid Y_{iu}, Z_{iu}) d\theta_{it} \notag \\
  &= 1 - \textrm{Pr}(\theta_{it} \in \Theta_1 \mid Y_{iu}, Z_{iu}),
\end{align}
which we expect to represent the probability of the subject $i$ being in the
luteal phase at time $t$.
Note that these probabilities are prospective when 
the conditional distribution of $\theta_{it}$ is a predictive distribution ($t > u$),
whereas they are retrospective when a smoothed distribution is used ($t < u$).

A subject $i$ can be judged as being in the first or second stage of the cycle 
based on the above probabilities.
We can decide $\theta_{it} \in \Theta_1$
when $\textrm{Pr}(\theta_{it} \in \Theta_1 \mid Y_{iu}, Z_{iu}) \geq 0.5$,
and $\theta_{it} \in \Theta_2$ otherwise.
We applied this decision rule in the analysis shown in section
\ref{sect:appl}, although other rules could be applied.

An R script illustrating how to implement the numerical procedures
for sequential Bayesian filtering, menstruation onset prediction, and judgement of stages
is available upon request.

\section{Application}
\label{sect:appl}

\subsection{Data}
\label{sect:data}
We organized data comprising daily recorded BBT and the day of menstrual onset, 
which were collected from a total of 3,784 women between 2007 and 2014
via a web service called {\it Ran's story} (QOL Corporation, Tomi, Japan).
Details of {\it Ran's story} service, of which users
were supposed to be ethnically Japanese, were described in
\cite{Fukaya2017}.
We focused on menstrual cycle data that were provided by users
aged between 15 and 54 years during this period, and
classified each menstrual cycle into eight age groups
(15--19, 20--24, 25--29, 30--34, 35--39, 40--44, 45--49, and 50--54 years)
based on the age of the user at the beginning of the cycle.

Cycles containing one or more BBT observations during the first seven days were defined
as applicable to our analyses, as we used BBT data in that time period
to standardize the level of BBT as explained below.
In addition, for each age group, the longest and shortest 5\% of cycles
were discarded to omit cycles with extreme length.
This data selection procedure resulted in a data set containing 25,622 cycles
provided by 3,533 unique users (Table \ref{table:data}, top rows).

In order to elucidate age-specific characteristics, we further generated a
subset of the above data set, which was randomly sampled while eliminating the
within-woman dependency as much as possible, and was used to estimate model parameters
and the accuracy of the prediction of menstruation
(Table \ref{table:data}, middle and bottom rows).
For each of the 20s and 30s age group, in which a large number of users
were registered, 450 users were selected randomly, so that 450 cycles can be sampled
from unique users.
Cycles were then assigned to 300 cycles for parameter estimation and
150 cycles for the assessment of predictive accuracy.
Similarly, for each of the 40s age group, 450 cycles were sampled randomly,
and then divided into 300 cycles for parameter estimation and 150 cycles
for the assessment of predictive accuracy.
However, due to the limited number of users,
not all cycles were attributed to unique users in these age groups.
Finally, for the late teens and early 50s age groups,
all available cycles were assigned randomly
for parameter estimation and the assessment of predictive accuracy, because
the number of cycles in these age groups was the most limited.
For the 15--19 years age group, 300 cycles were used for parameter estimation
and the remaining 111 cycles
were used for the assessment of predictive accuracy.
For the 50--54 years age group, 120 cycles were assigned for parameter estimation
and the remaining 52 cycles
were used for the assessment of predictive accuracy.

As the level of BBT time series varies between cycles,
we used standardized BBT data in the following analyses.
For each cycle, BBT time series was standardized by subtracting the
median of BBT recorded in the first seven days of the cycle from the raw BBT data.

\subsection{Parameter estimation}

For each age group, an explicit biphasic menstrual cycle model was
fitted to data for parameter estimation (Table \ref{table:data}, middle rows).
We treated cycles within each age group as independent, and assumed that
a set of parameters specific to the age group apply to them.
In order to fit the model and evaluate log-likelihood, we used a non-Gaussian filter
discretizing the state space into 512 intervals.

\begin{figure*}[p]
\begin{center}
  \includegraphics[bb=0 0 1080 504,width=6.5in]{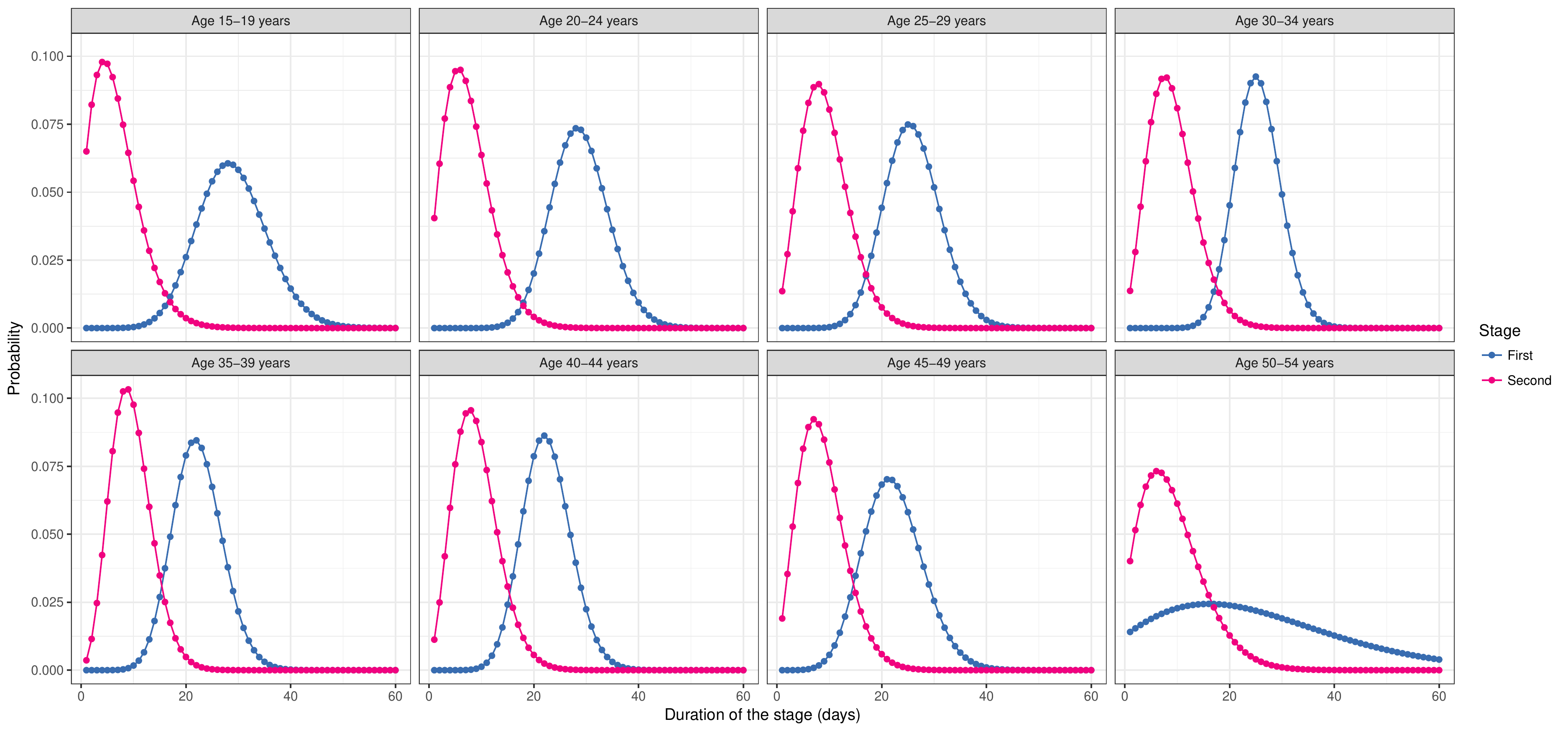}
\end{center}
\caption{
  Distribution of the stage length derived from estimated parameters. 
  The probability distribution of the length of $m$-th stage $f_m(k) (m = 1, 2)$ was obtained as: $f_m(1) = 1 - G(0.5; \alpha_{m}, \beta_{m})$ and for $k > 1$, $f_m(k) = G\left\{0.5; (k-1)\alpha_{m}, \beta_{m}\right\} - G(0.5; k\alpha_{m}, \beta_{m})$, where $G(\hspace{.1em}\cdot\hspace{.1em} ; s, r)$ is the distribution function of the gamma distribution with shape parameter $s$ and rate parameter $r$.
  Each panel depicts the results for an age group.
}
\label{fig:PhaseDuration}
\end{figure*}

Parameter values were distinctly different between two stages (Table \ref{table:param}).
System model parameters were characterized by larger values of
$\alpha$ and $\beta$ in the first stage, except for the oldest age group (50--54 years).
These estimates generally implied that the first stage is
on average longer and is more variable than the second stage
(Figure \ref{fig:PhaseDuration}).
Between-age differences in the stage length distribution were
also discernible: the stage length tended to become variable in late teens
and early 50s (Figure \ref{fig:PhaseDuration}).

As expected, in all age groups, mean BBT was estimated to be higher
in the second stage: the increase in BBT $(\mu_2 - \mu_1)$ ranged from
0.342 to 0.413 (Table \ref{table:param}).
No clear difference was found in the standard deviation of observation errors
of BBT (Table \ref{table:param}).

\subsection{Accuracy of the prediction of menstruation onset}

The accuracy of the prediction for the day of menstruation onset of the
explicit biphasic menstrual cycle model was compared to that of several variants of
state-space models for menstrual cycles and the conventional calendar calculation method.
Models used for this comparison are summarized in Table \ref{table:models}.
Like the fully explicit model, parameters of other state-space models were estimated by
fitting the model to data for parameter estimation (Table \ref{table:data}, middle rows).
The predictive error was measured by the root mean square error (RMSE).
As the explicit model and implicit model can adaptively adjust the prediction
based on the daily BBT records, for these models, RMSE was estimated for
several points in time within the cycle; namely, at the day of onset of the
previous menstruation, as well as 21, 14, 7, 6, 5, 4, 3, 2, and 1 day(s) before
the day of the next menstruation onset.
Cycles that were shorter than 21 days were omitted from the RMSE calculation for
21 days before the day of onset of next menstruation.
The calendar calculation method predicts the next menstruation day as
the day after a fixed number of days from the onset of preceding menstruation,
which thus does not update the prediction within the cycle.
For each age class, we used a fixed number of days for calendar-based prediction
that gave the lowest RMSE.

Results are shown in Figure \ref{fig:PredErr}.
Overall, the predictive error (RMSE) was found to be larger in
young age groups (late teens and early 20s) and the oldest age group.
In the fully explicit model, RMSE was in general gradually decreased as the
onset day of next menstruation approached.
In all age groups, the model prediction was superior to the conventional
calendar calculation method in the last few days of the cycle.
On the other hand, RMSE tended to increase in the late stage of the cycle
in the restricted explicit model, except for the oldest age group.
The model prediction may be considerably worse than the calendar calculation method,
especially in the middle age classes.
Implicit models tended to give relatively poor prediction in the early stages of the cycle.
However, in the last few days of the cycle, they attained small RMSE that was comparable
to the fully explicit model.
In the 50--54 years age group, 
results from the implicit models were identical.
This was because in this age group, the system model parameters (i.e., $\alpha$ and $\beta$)
were estimated to be very small, which almost always predict the onset of next menstruation
occurring in the following day (a phenomena that was previously known to occur in the
implicit models; \citealt{Fukaya2017}).
In this setting, RMSE decreases constantly, and finally reaches to zero 1 day before
menstruation onset.
However, such a manner of prediction is of course meaningless.
The results together indicate that introducing the biphasic structure into
the system model was critical to improve predictions in a wide range of age class.

\subsection{Distribution of the length of two stages}

\begin{figure*}[p]
\begin{center}
  \includegraphics[bb=0 0 1080 648,width=6.5in]{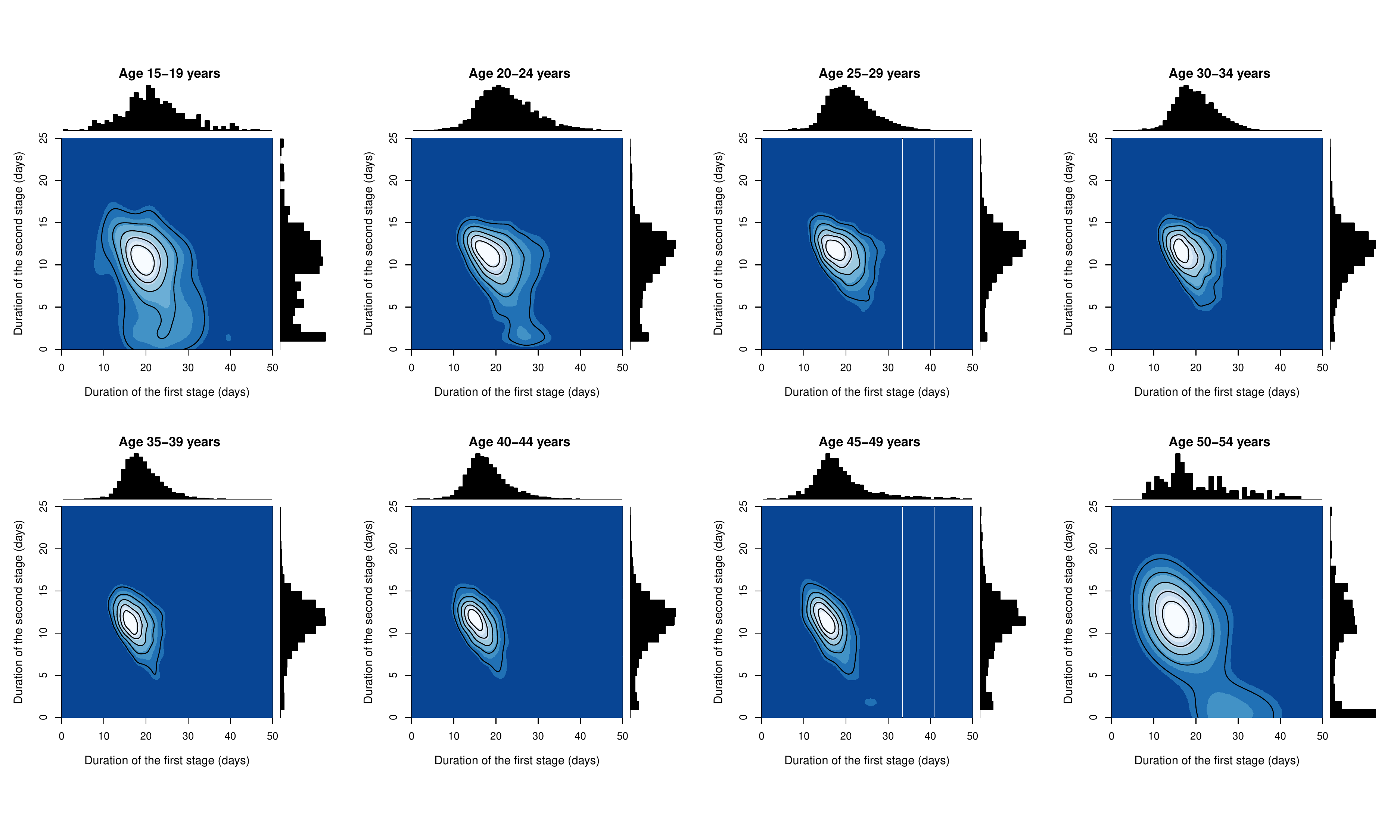}
\end{center}
\caption{
  The contour plots show the kernel density estimates of the durations of the first and second stages. Dense regions are drawn in brighter colors. The marginal distributions of the stage lengths are shown on the upper and right side of the plots. 
}
\label{fig:correlation}
\end{figure*}

Based on the stage identification method described in
section \ref{sect:phase}, we determined lengths of the first and second stages of
each menstrual cycle in the entire data set (Table \ref{table:data}, top rows).
We used the smoothed probability distribution of the menstrual phase
to determine the conditional probability of each stage.
The joint and marginal distributions of the length of those stages are shown
in Figure \ref{fig:correlation}.
In all age groups, there was a negative correlation between
the length of the first and second stages.
We also found that in the second stage of some age groups,
there was a peak located at 0 or 1, indicating the existence of monophasic cycles.

The summary of two stage lengths is shown in Table \ref{table:stage_summary}.
We found that in all age groups, the first stage is longer,
on average, and is more variable than the second stage.
The standard deviation of both stages tended to increase in either end of the age group.
Furthermore, the percentage of monophasic cycles, which were arbitrarily defined as
cycles that the length of the second stage was estimated to be less than three days,
was also higher in these age groups.

\section{Discussion}
\label{sect:disc}

In the present study, we developed a self-excited autoregressive state-space model
that explicitly accommodated the biphasic nature of the menstrual cycle,
as an extension of the state-space model for the menstrual cycle
proposed by \cite{Fukaya2017}.
The present model was fitted to menstrual cycle data obtained from 
a large number of women in different age groups. We
found that the estimated parameters were clearly different between
the first and second stages of the cycle (Table \ref{table:param}).
Mean BBT was estimated to be lower in the first stage of the cycle
and to increase by approximately $0.4\ {}^\circ\mathrm{C}$ in the second stage of the cycle.
This result is consistent with those of \cite{Scarpa2009},
who analyzed BBT data obtained from a European fertility study demonstrating
that, in normal menstrual cycles exhibiting a biphasic pattern,
the BBT shifts an average of $0.4\ {}^\circ\mathrm{C}$.
Furthermore, estimates of the system model parameters suggested that
the model predicts the length of the first stage to be
longer and more variable than that of the second stage (Figure \ref{fig:PhaseDuration}).
This result coincides with the fact that 
variability in the MCL can be mainly
attributed to variability in the length of the follicular phase \citep{Fehring2006}.
We therefore conclude that the proposed model adequately 
captured the characteristics of the two phases in the ovarian cycle
(i.e., the follicular and luteal phases),
based on time series data of the BBT and menstruation.

\begin{figure*}[p]
\begin{center}
  \includegraphics[bb=0 0 1080 720,width=6.5in]{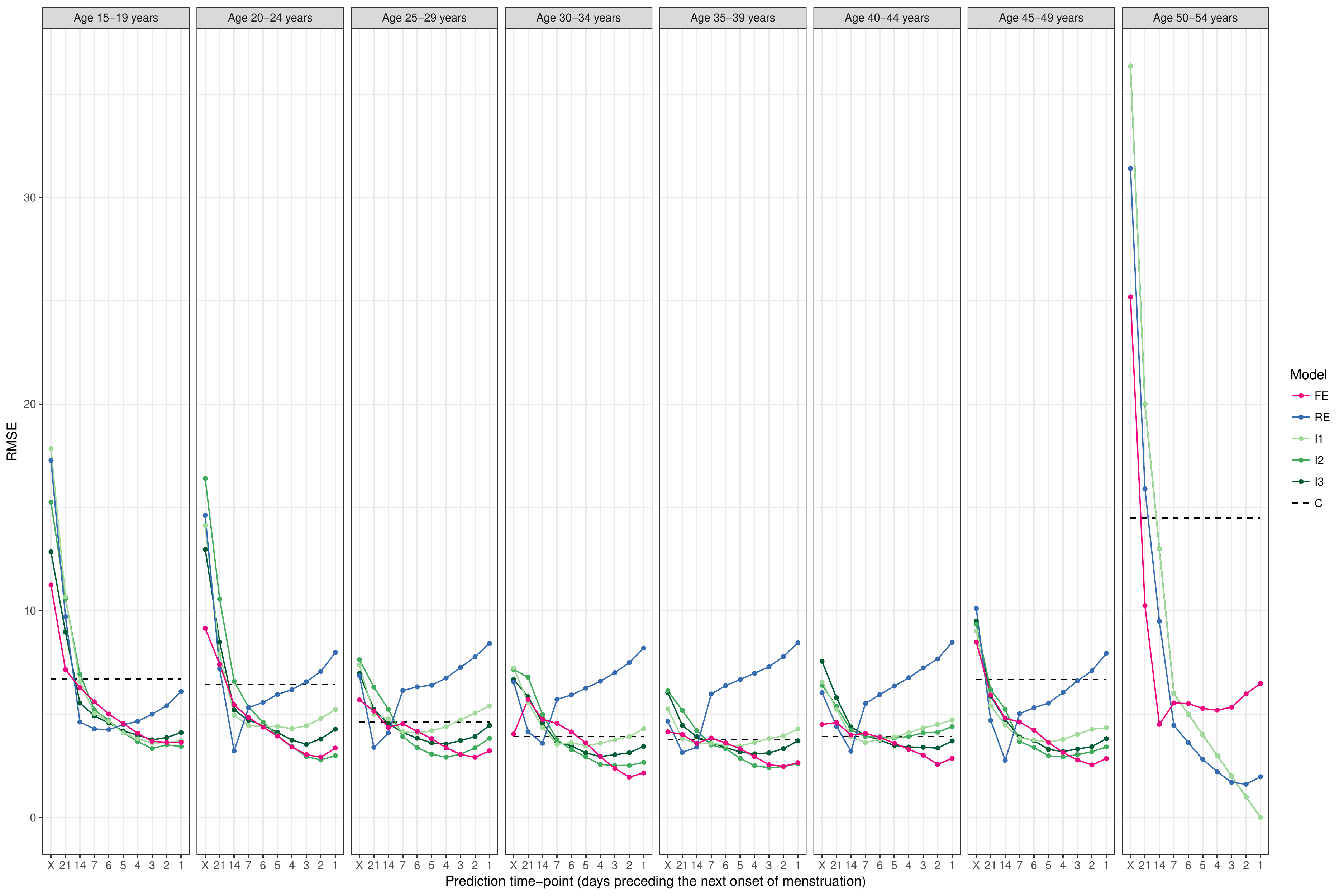}
\end{center}
\caption{
  Root mean square error (RMSE) of prediction of the day of onset of menstruation across the day at which prediction was made. ``X'', on the horizontal axis, indicates the day of onset of previous menstruation.
FE: Fully explicit model; RE: Restricted explicit model; I1: Implicit model 1; I2: Implicit model 2; I3: Implicit model 3; C: Calendar calculation method; refer to Table \ref{table:models} for details.
}
\label{fig:PredErr}
\end{figure*}

Since the proposed model is an extension of the state-space model proposed by
\cite{Fukaya2017},
it can be used to obtain a predictive distribution of the length of the
current menstrual cycle sequentially.
Our analysis showed that with the accumulation of the
within-cycle trajectory of the BBT, the proposed model can provide a prediction
that was superior to that for the conventional calendar calculation (Figure \ref{fig:PredErr}).
Furthermore, an explicit consideration of the biphasic characteristics enabled the
proposed model to give a better prediction in a wide range of age groups compared to
the implicit models (Figure \ref{fig:PredErr}).
We note that \cite{Bortot2010} proposed another state-space modeling framework
that can be used to predict the length of the current menstrual cycle,
based on the subject's past time series of the MCL.
We did not compare the predictive accuracy of the proposed model to that of the 
model of \cite{Bortot2010}, 
because the data in the present study did not include a sufficiently long time series
for each subject.
A comparison between the models of \cite{Bortot2010} and
\cite{Fukaya2017} is reported elsewhere \citep{Fukaya2017}.

Using smoothed probabilities, we determined the length of the first and second stages
of each examined menstrual cycle, 
clarifying the statistical characteristics of the length of each stage 
in different age groups (Figure \ref{fig:correlation}, Table \ref{table:stage_summary}).
\cite{Fehring2006} collected data on the length of the
follicular and luteal phases in 165 women aged 21--44 years.
They reported that mean, median, mode, and standard deviation
of the phase length was
16.5, 16, 15, and 3.4 days for follicular phase, and 
12.4, 13, 13, and 2.0 days for luteal phase, respectively \citep{Fehring2006}.
Although the data we examined was collected in a more opportunistic manner,
we found a fairly close central tendency in the distribution of the length of the first and
second stages in each age group, especially when monophasic cycles were excluded
(Table \ref{table:stage_summary}).
Compared to the previous study, however, variation was larger in our data set.
A possible explanation for this is that 
the range of MCL (Table \ref{table:data})
was wider in our data set than the previous study, in which 
the MCL recorded was limited between 21 and 42 days \citep{Fehring2006}.
Another reason may be an inclusion of monophasic cycles in the data set.
In contrast to \cite{Fehring2006}, where cycles indicating no sign of
LH surge were omitted, monophasic cycles were not ruled out beforehand in our analysis.
Even when the subset of data in which cycles that had estimated length of the
second stage as less than three days were removed (Table \ref{table:stage_summary}, middle rows),
a number of monophasic cycles in which the length of the
second stage was very short may remain.

Using a large data set of menstrual cycles,
our assessment of the stage length distribution also enabled us to find a
mild negative correlation between the lengths of the first and
second stages, which has been previously reported \citep{Fehring2006},
in a wide range of age groups.
Furthermore, we found that the length of the second stage was extremely short
in a portion of the menstrual cycles, indicating the existence of
menstrual cycles exhibiting a monophasic BBT pattern, which has been
recognized to occur \citep{Barron2005}.
These results affirm that the explicit biphasic menstrual cycle model can provide a reasonable
judgement of the subject being in the follicular or luteal phase.
We note that our analysis highlighted the among-age group difference in the
phase length distribution, where the durations of follicular and luteal phase
were more variable in the late teens and early 50s (Table \ref{table:stage_summary}).
It was also evident that monophasic cycles appear most often
in these age groups (Table \ref{table:stage_summary}).

As long as the model adequately distinguishes the follicular and luteal phases,
the conditional probabilities for a subject
being in the first or second stages of the menstrual cycle
could be used for a model-based judgement for the day of ovulation.
Several methods have been proposed to objectively identify the day of ovulation based on the BBT,
which include a widely used rule of thumb called the three over six rule 
\citep{Bortot2010,Colombo2000,Bigelow2004},
a method based on the cumulative sum test \citep{Royston1980}, and
a stopping rule based on a change-point model \citep{Carter1981}.
A limitation shared among these methods is that
they may be difficult to apply, or less effective even if used \citep{Carter1981},
when observations in the BBT time series are missing.
However, missing values can be formally handled
in the state-space modeling framework; thus, the proposed method 
does not suffer from missing observations.
Furthermore, judgements regarding ovulation can be made in a prospective,
real-time, or retrospective manner, depending on the type of conditional distribution
(i.e., predictive, filtering, or smoothed distribution, respectively).
The proposed model may therefore be considered as a new approach
to identify the day of ovulation based on the BBT,
for which previous methods are considered to be error-prone
\citep{Barron2005, Scarpa2009}.
However, the magnitude of the identification error is currently unknown,
and further validation is required in the future.

As described in Section \ref{sect:ssm}, the model can accommodate variations in the parameters 
by including covariates,
which can be useful in explaining differences in the characteristics of the menstrual cycles 
associated with subject-specific characteristics and/or
conditions that vary between cycles within a subject \citep{Murphy1995,Liu2004}.
Another possibility for modeling variability in the parameters involves the
inclusion of random effects.
However, this complicates the calculation of the log-likelihood considerably,
rendering parameter estimation more challenging.
Specifically, the inclusion of random effects requires an alternative
estimation approach, such as Bayesian estimation using the Markov chain Monte Carlo method,
which may result in a considerable increase in computational time.
Finally, we note that the proposed method assumes that the BBT of the subject
fluctuates under natural conditions.
Therefore, the proposed method may not be useful for women utilizing hormonal contraception,
which can interfere with physiological phenomena related to the menstrual cycle.

\section*{Acknowledgements}

We are grateful to K. Shimizu and T. Matsui
who provided valuable comments on this research.
This work was supported by the ``Research and Development on Fundamental
and Utilization Technologies for Social Big Data'', as Commissioned Research of the National
Institute of Information and Communications Technology (NICT), Japan (No 178A02).
This research was initiated by an application to the ISM
Research Collaboration Start-up program (No RCSU2013--08) from M. Kitazawa.
This research was supported by an allocation of computing resources of the
SGI ICE X and SGI UV 2000 supercomputers from the Institute of Statistical Mathematics.


\appendix

\section{Conditional probabilities for the day of menstruation onset}
\label{sect:appa}

In the following, we use mathematical notations listed in Table \ref{table:app},
in addition to those described in Section \ref{sect:model}.
For $\theta_{it} \in \Theta_2$, $f(k \mid \theta_{it})$ can be obtained
straightforwardly by using the distribution function of gamma distribution.
For $k = 1$,
\begin{align}
  f(1 \mid \theta_{it}) 
  &= \textrm{Pr}(\epsilon^{(2)}_{i,t+1} \geq \thllh) \notag \\
  &= \int_{\thllh}^{\infty} g(x; \alpha_{i2}, \beta_{i2}) dx \notag \\
  &= 1 - G(\thllh; \alpha_{i2}, \beta_{i2}),
\end{align}
\noindent{where} $\lceil x \rceil$ is the ceiling function that returns the
smallest following integer for $x$.
For \noindent{$k > 1$},
\begin{align}
f(k \mid \theta_{it}) 
&= \textrm{Pr}\left(\sum_{r=1}^k \epsilon^{(2)}_{i,t+r} \geq \thllh \right) 
- \textrm{Pr}\left(\sum_{r=1}^{k-1} \epsilon^{(2)}_{i,t+r} \geq \thllh \right) \notag \\
    &= \int_{\thllh}^{\infty} 
    g(x; k\alpha_{i2}, \beta_{i2}) dx - \int_{\thllh}^{\infty} 
    g(x; (k - 1) \alpha_{i2}, \beta_{i2}) dx \notag  \\
    &= G\left\{\thllh; (k-1)\alpha_{i2}, \beta_{i2}\right\} 
  - G(\thllh; k\alpha_{i2}, \beta_{i2}).
\end{align}
\noindent{Note} that these equations are analogous to the derivation of the
conditional probability under the implicit model \citep{Fukaya2017}.

For $\theta_{it} \in \Theta_1$, the calculation of $f(k \mid \theta_{it})$ is more complicated
because system model parameters should switch at a particular, but unknown, point of time.
We define $\phi(j,k \mid \theta_{it})$ ($j = 1, \dots, k$) as the probability that 
the first stage of the cycle lasts until the $t + j -1$th day
and the next menstruation occurs on the $t + k$th day.
$f(k \mid \theta_{it})$ is then expressed as
\begin{equation}
  f(k \mid \theta_{it}) = \sum_{j=1}^k \phi(j,k \mid \theta_{it}),
\end{equation}
\noindent{where} $\phi(j, k \mid \theta_{it})$ is calculated as follows.

\begin{itemize}
  \item For $j=1$ and $k=1$,
\end{itemize}
\begin{align}
  \phi(1, 1 \mid \theta_{it}) &= \textrm{Pr}(\epsilon^{(1)}_{i,t+1} \geq \thllh) \notag \\
  &= \int_{\thllh}^{\infty} g(x; \alpha_{i1}, \beta_{i1}) dx \notag \\
  &= 1 - G(\thllh; \alpha_{i1}, \beta_{i1}).
\end{align}

\begin{itemize}
  \item For $j=1$ and $k=2$,
\end{itemize}
\begin{align}
  &\phi(1, 2 \mid \theta_{it}) \notag \\
  &= \textrm{Pr} \bigl\{ \underbrace{\thlfh \leq \epsilon^{(1)}_{i,t+1} <
  \thllh}_{A} \bigr\} \times
  \textrm{Pr}\left\{\epsilon^{(1)}_{i,t+1} + \epsilon^{(2)}_{i,t+2} \geq \thllh \mid A \right\} \notag \\
  &= \int_{\thlfh}^{\thllh} \pi_{v_{i1}}(x)dx 
  \times \int_{\thllh}^{\infty} \left[ \pi^*_{v_{i1}}\left\{\thlfh, \thllh\right\} * g(\alpha_{i2}, \beta_{i2}) \right](x)dx \notag \\
  &= \left[ G\left( \thllh; \alpha_{i1}, \beta_{i1} \right)
  - G\left\{\thlfh; \alpha_{i1}, \beta_{i1} \right\} \right] \notag \\ 
  &\hspace{2em} \times \int_{\thllh}^{\infty} \left[ \pi^*_{v_{i1}}\left\{\thlfh, \thllh\right\} * g(\alpha_{i2}, \beta_{i2}) \right](x)dx.
\end{align}

\begin{itemize}
  \item For $j=1$ and $k>2$,
\end{itemize}
\begin{align}
  &\phi(1, k \mid \theta_{it}) \notag \\
  &= \textrm{Pr}\bigl\{ \underbrace{\thlfh \leq \epsilon^{(1)}_{i,t+1} <
  \thllh}_{A} \bigr\} \notag \\
  &\hspace{2em} \times \textrm{Pr}\Biggl\{ \underbrace{\thlfh \leq \epsilon^{(1)}_{i,t+1} + \sum_{r=2}^{k-1} \epsilon^{(2)}_{i,t+r} < \thllh}_{B} \mid A \Biggr\} \notag  \\
  &\hspace{2em} \times \textrm{Pr}\left\{\epsilon^{(1)}_{i,t+1} + \sum_{r=2}^{k-1} \epsilon^{(2)}_{i,t+r} + \epsilon^{(2)}_{i,t+k} \geq \thllh \mid A, B \right\}  \notag \\
  &= \int_{\thlfh}^{\thllh} \pi_{v_{i1}}(x)dx 
  \times \int_{\thlfh}^{\thllh} \pi_{w_{i1k}}(x)dx  \notag \\
  &\hspace{2em} \times \int_{\thllh}^{\infty} \left[ \pi^*_{w_{i1k}}\left\{\thlfh, \thllh\right\} * g\left(\alpha_{i2}, \beta_{i2}\right) \right](x)dx \notag \\
  &= \left[ G\left( \thllh; \alpha_{i1}, \beta_{i1} \right)
  - G\left\{\thlfh; \alpha_{i1}, \beta_{1} \right\} \right] \notag \\
  &\hspace{2em} \times \int_{\thlfh}^{\thllh} \left[ \pi^*_{v_{i1}}\left\{\thlfh, \thllh\right\} * 
  g\left\{ (k-2)\alpha_{i2}, \beta_{i2} \right\} \right](x)dx \notag  \\
  &\hspace{2em} \times \int_{\thllh}^{\infty} \left[ \pi^*_{w_{i1k}}\left\{\thlfh, \thllh\right\} * g\left(\alpha_{i2}, \beta_{i2}\right) \right](x)dx.
\end{align}

\begin{itemize}
  \item For $j>1$ and $k=j$,
\end{itemize}
\begin{align}
  &\phi(j, k \mid \theta_{it})  \notag \\
  &= \textrm{Pr} \Biggl\{ \underbrace{\sum_{r=1}^{j-1} \epsilon^{(1)}_{i,t+r} < \thlfh}_{A} \Biggr\} 
  \times \textrm{Pr}\left\{\sum_{r=1}^{j-1} \epsilon^{(1)}_{i,t+r} + \epsilon^{(1)}_{i,t+k} \geq \thllh \mid A \right\} \notag \\
  &= \int_0^{\thlfh} \pi_{u_{ij}}(x)dx 
  \times \int_{\thllh}^{\infty} \pi_{v_{ij}}(x) dx \notag \\
  &= G\left\{ \thlfh; (j-1)\alpha_{i1}, \beta_{i1} \right\} \notag \\ 
  &\hspace{2em} \times \int_{\thllh}^{\infty} \left[ \pi^*_{u_{ij}}\left\{0, \thlfh\right\} * g(\alpha_{i1}, \beta_{i1}) \right](x) dx.
\end{align}

\begin{itemize}
  \item For $j>1$ and $k=j+1$,
\end{itemize}
\begin{align}
  &\phi(j, k \mid \theta_{it}) \notag \\
  &= \textrm{Pr} \Biggl\{ \underbrace{\sum_{r=1}^{j-1} \epsilon^{(1)}_{i,t+r} < \thlfh}_{A} \Biggr\} \notag \\
  &\hspace{2em} \times \textrm{Pr}\Biggl\{\underbrace{\thlfh \leq \sum_{r=1}^{j-1} \epsilon^{(1)}_{i,t+r} + \epsilon^{(1)}_{i,t+j} < \thllh}_{B} \mid A \Biggr\}  \notag \\
  &\hspace{2em} \times \textrm{Pr}\left\{\sum_{r=1}^{j-1} \epsilon^{(1)}_{i,t+r} + \epsilon^{(1)}_{i,t+j} + \epsilon^{(2)}_{i,t+k} \geq \thllh \mid A, B \right\} \notag  \\
  &= \int_0^{\thlfh} \pi_{u_{ij}}(x)dx 
  \times \int_{\thlfh}^{\thllh} \pi_{v_{ij}}(x)dx \notag  \\
  &\hspace{2em} \times \int_{\thllh}^{\infty} \left[ \pi^*_{v_{ij}}\left\{\thlfh, \thllh\right\} * g(\alpha_{i2}, \beta_{i2}) \right](x)dx \notag \\
  &= G\left\{ \thlfh; (j-1)\alpha_{i1}, \beta_{i1} \right\} \notag \\
  &\hspace{2em} \times \int_{\thlfh}^{\thllh} \left[ \pi^*_{u_{ij}}\left\{ 0, \thlfh \right\} * g(\alpha_{i1}, \beta_{i1}) \right](x)dx  \notag \\
  &\hspace{2em} \times \int_{\thllh}^{\infty} \left[ \pi^*_{v_{ij}}\left\{\thlfh, \thllh\right\} * g(\alpha_{i2}, \beta_{i2}) \right](x)dx. 
\end{align}

\begin{itemize}
  \item For $j>1$ and $k>j+1$,
\end{itemize}
\begin{align}
  &\phi(j, k \mid \theta_{it}) \notag \\
  &= \textrm{Pr} \Biggl\{ \underbrace{\sum_{r=1}^{j-1} \epsilon^{(1)}_{i,t+r} < \thlfh}_{A} \Biggr\} \notag \\
  &\hspace{2em} \times \textrm{Pr}\Biggl\{\underbrace{\thlfh \leq \sum_{r=1}^{j-1} \epsilon^{(1)}_{i,t+r} + \epsilon^{(1)}_{i,t+j} < \thllh}_{B} \mid A \Biggr\} \notag  \\
  &\hspace{2em} \times \textrm{Pr}\Biggl\{\underbrace{\thlfh \leq \sum_{r=1}^{j-1} \epsilon^{(1)}_{i,t+r} + \epsilon^{(1)}_{i,t+j} + \sum_{r=j+1}^{k-1}\epsilon^{(2)}_{i,t+r} < \thllh}_{C} \mid A, B \Biggr\} \notag  \\
  &\hspace{2em} \times \textrm{Pr}\left\{\sum_{r=1}^{j-1} \epsilon^{(1)}_{i,t+r} + \epsilon^{(1)}_{i,t+j} + \sum_{r=j+1}^{k-1}\epsilon^{(2)}_{i,t+r} + \epsilon^{(2)}_{i,t+k} \geq \thllh \mid A, B, C \right\}  \notag \\
  &= \int_0^{\thlfh} \pi_{u_{ij}}(x)dx 
  \times \int_{\thlfh}^{\thllh} \pi_{v_{ij}}(x)dx  
    \times \int_{\thlfh}^{\thllh} \pi_{w_{ijk}}(x)dx  \notag \\
    &\hspace{2em} \times \int_{\thllh}^{\infty} \left[ \pi^*_{w_{ijk}}\left\{\thlfh, \thllh\right\} * g(\alpha_{i2}, \beta_{i2}) \right] (x)dx \notag  \\
    &= G\left\{ \thlfh; (j-1)\alpha_{i1}, \beta_{i1} \right\} \notag \\
    &\hspace{2em} \times \int_{\thlfh}^{\thllh} \left[ \pi^*_{u_{ij}}\left\{ 0, \thlfh \right\} * g(\alpha_{i1}, \beta_{i1}) \right](x)dx  \notag \\
    &\hspace{2em} \times \int_{\thlfh}^{\thllh} \left[ \pi^*_{v_{ij}}\left\{ \thlfh, \thllh \right\} * g\left\{ (k-j-1)\alpha_{i2}, \beta_{i2} \right\} \right](x)dx  \notag \\
    &\hspace{2em} \times \int_{\thllh}^{\infty} \left[ \pi^*_{w_{ijk}}\left\{\thlfh, \thllh\right\} * g(\alpha_{i2}, \beta_{i2}) \right] (x)dx.
\end{align}

Convolutions of probability density distributions appear in above equations
does not have a closed-form solution, and thus needed to be evaluated numerically.
In our R implementation (available upon request),
we used \texttt{distr} package \citep{Ruckdeschel2014}
which provides convolution algorithm based on the fast Fourier transform (FFT).


\begin{thebibliography}{}

\bibitem[\protect\citeauthoryear{Barrett and Marshall}{Barrett and
  Marshall}{1969}]{Barrett1969}
Barrett, J.~C. and Marshall, J. (1969).
\newblock The risk of conception on different days of the menstrual cycle.
\newblock {\em Population Studies} {\bf 23,} 455--461.

\bibitem[\protect\citeauthoryear{Barron and Fehring}{Barron and
  Fehring}{2005}]{Barron2005}
Barron, M.~L. and Fehring, R.~J. (2005).
\newblock {Basal body temperature assessment: is it useful to couples seeking
  pregnancy?}
\newblock {\em The American Journal of Maternal Child Nursing} {\bf 30,}
  290--296.

\bibitem[\protect\citeauthoryear{Bigelow, Dunson, Stanford, Ecochard, Gnoth,
  and Colombo}{Bigelow et~al.}{2004}]{Bigelow2004}
Bigelow, J.~L., Dunson, D.~B., Stanford, J.~B., Ecochard, R., Gnoth, C., and
  Colombo, B. (2004).
\newblock Mucus observations in the fertile window: a better predictor of
  conception than timing of intercourse.
\newblock {\em Human Reproduction} {\bf 19,} 889--892.

\bibitem[\protect\citeauthoryear{Bortot, Masarotto, and Scarpa}{Bortot
  et~al.}{2010}]{Bortot2010}
Bortot, P., Masarotto, G., and Scarpa, B. (2010).
\newblock {Sequential predictions of menstrual cycle lengths}.
\newblock {\em Biostatistics} {\bf 11,} 741--755.

\bibitem[\protect\citeauthoryear{Carter and Blight}{Carter and
  Blight}{1981}]{Carter1981}
Carter, R.~L. and Blight, B.~J. (1981).
\newblock A Bayesian change-point problem with an application to the prediction
  and detection of ovulation in women.
\newblock {\em Biometrics} {\bf 37,} 743--751.

\bibitem[\protect\citeauthoryear{Colombo and Masarotto}{Colombo and
  Masarotto}{2000}]{Colombo2000}
Colombo, B. and Masarotto, G. (2000).
\newblock Daily fecundability: first results from a new data base.
\newblock {\em Demographic research} {\bf 3,} 5.

\bibitem[\protect\citeauthoryear{Dunson, Baird, Wilcox, and Weinberg}{Dunson
  et~al.}{1999}]{Dunson1999}
Dunson, D.~B., Baird, D.~D., Wilcox, A.~J., and Weinberg, C.~R. (1999).
\newblock Day-specific probabilities of clinical pregnancy based on two studies
  with imperfect measures of ovulation.
\newblock {\em Human Reproduction} {\bf 14,} 1835--1839.

\bibitem[\protect\citeauthoryear{Dunson, Colombo, and Baird}{Dunson
  et~al.}{2002}]{Dunson2002}
Dunson, D.~B., Colombo, B., and Baird, D.~D. (2002).
\newblock Changes with age in the level and duration of fertility in the
  menstrual cycle.
\newblock {\em Human Reproduction} {\bf 17,} 1399--1403.

\bibitem[\protect\citeauthoryear{Dunson and Weinberg}{Dunson and
  Weinberg}{2000}]{Dunson2000}
Dunson, D.~B. and Weinberg, C.~R. (2000).
\newblock Modeling human fertility in the presence of measurement error.
\newblock {\em Biometrics} {\bf 56,} 288--292.

\bibitem[\protect\citeauthoryear{Dunson, Weinberg, Baird, Kesner, and
  Wilcox}{Dunson et~al.}{2001}]{Dunson2001}
Dunson, D.~B., Weinberg, C.~R., Baird, D.~D., Kesner, J.~S., and Wilcox, A.~J.
  (2001).
\newblock Assessing human fertility using several markers of ovulation.
\newblock {\em Statistics in Medicine} {\bf 20,} 965--978.

\bibitem[\protect\citeauthoryear{Ecochard}{Ecochard}{2006}]{Ecochard2006}
Ecochard, R. (2006).
\newblock Heterogeneity in fecundability studies: issues and modelling.
\newblock {\em Statistical Methods in Medical Research} {\bf 15,} 141--160.

\bibitem[\protect\citeauthoryear{Fehring, Schneider, and Raviele}{Fehring
  et~al.}{2006}]{Fehring2006}
Fehring, R.~J., Schneider, M., and Raviele, K. (2006).
\newblock Variability in the phases of the menstrual cycle.
\newblock {\em Journal of Obstetric, Gynecologic, and Neonatal Nursing} {\bf
  35,} 376--384.

\bibitem[\protect\citeauthoryear{Fukaya, Kawamori, Osada, Kitazawa, and
  Ishiguro}{Fukaya et~al.}{2017}]{Fukaya2017}
Fukaya, K., Kawamori, A., Osada, Y., Kitazawa, M., and Ishiguro, M. (2017).
\newblock The forecasting of menstruation based on a state-space modeling of
  basal body temperature time series.
  \newblock {\em Statistics in Medicine} {(Published online, doi:
10.1002/sim.7345),} 1--19.

\bibitem[\protect\citeauthoryear{Guo, Manatunga, Chen, and Marcus}{Guo
  et~al.}{2006}]{Guo2006}
Guo, Y., Manatunga, A.~K., Chen, S., and Marcus, M. (2006).
\newblock {Modeling menstrual cycle length using a mixture distribution}.
\newblock {\em Biostatistics} {\bf 7,} 100--114.

\bibitem[\protect\citeauthoryear{Harlow and Zeger}{Harlow and
  Zeger}{1991}]{Harlow1991}
Harlow, S.~D. and Zeger, S.~L. (1991).
\newblock {An application of longitudinal methods to the analysis of menstrual
  diary data}.
\newblock {\em Journal of Clinical Epidemiology} {\bf 44,} 1015--1025.

\bibitem[\protect\citeauthoryear{Huang, Elliott, and Harlow}{Huang
  et~al.}{2014}]{Huang2014}
Huang, X., Elliott, M.~R., and Harlow, S.~D. (2014).
\newblock {Modelling menstrual cycle length and variability at the approach of
  menopause by using hierarchical change point models}.
\newblock {\em Journal of the Royal Statistical Society: Series C (Applied
  Statistics)} {\bf 63,} 445--466.

\bibitem[\protect\citeauthoryear{Ives and Dakos}{Ives and
  Dakos}{2012}]{Ives2012}
Ives, A.~R. and Dakos, V. (2012).
\newblock Detecting dynamical changes in nonlinear time series using locally
  linear state-space models.
\newblock {\em Ecosphere} {\bf 3,} 1--15.

\bibitem[\protect\citeauthoryear{Kitagawa}{Kitagawa}{1987}]{Kitagawa1987}
Kitagawa, G. (1987).
\newblock {Non-Gaussian state-space modeling of nonstationary time series}.
\newblock {\em Journal of the American Statistical Association} {\bf 82,}
  1032--1041.

\bibitem[\protect\citeauthoryear{Kitagawa}{Kitagawa}{2010}]{Kitagawa2010}
Kitagawa, G. (2010).
\newblock {\em {Introduction to Time Series Modeling}}.
\newblock Chapman \& Hall/CRC, Boca Raton.

\bibitem[\protect\citeauthoryear{Lin, Raz, and Harlow}{Lin
  et~al.}{1997}]{Lin1997}
Lin, X., Raz, J., and Harlow, S.~D. (1997).
\newblock Linear mixed models with heterogeneous within-cluster variances.
\newblock {\em Biometrics} {\bf 53,} 910--923.

\bibitem[\protect\citeauthoryear{Liu, Manatunga, Peng, and Marcus}{Liu
  et~al.}{2017}]{Liu2017}
Liu, S., Manatunga, A.~K., Peng, L., and Marcus, M. (2017).
\newblock A joint modeling approach for multivariate survival data with random
  length.
\newblock {\em Biometrics} {\bf 73,} 666--677.

\bibitem[\protect\citeauthoryear{Liu, Gold, Lasley, and Johnson}{Liu
  et~al.}{2004}]{Liu2004}
Liu, Y., Gold, E.~B., Lasley, B.~L., and Johnson, W.~O. (2004).
\newblock Factors affecting menstrual cycle characteristics.
\newblock {\em American Journal of Epidemiology} {\bf 160,} 131--140.

\bibitem[\protect\citeauthoryear{Lum, Sundaram, Louis, and Louis}{Lum
  et~al.}{2016}]{Lum2016}
Lum, K.~J., Sundaram, R., Louis, G. M.~B., and Louis, T.~A. (2016).
\newblock A Bayesian joint model of menstrual cycle length and fecundity.
\newblock {\em Biometrics} {\bf 72,} 193--203.

\bibitem[\protect\citeauthoryear{McLain, Lum, and Sundaram}{McLain
  et~al.}{2012}]{McLain2012}
McLain, A.~C., Lum, K.~J., and Sundaram, R. (2012).
\newblock A joint mixed effects dispersion model for menstrual cycle length and
  time-to-pregnancy.
\newblock {\em Biometrics} {\bf 68,} 648--656.

\bibitem[\protect\citeauthoryear{Murphy, Bentley, and O'Hanesian}{Murphy
  et~al.}{1995}]{Murphy1995}
Murphy, S.~A., Bentley, G.~R., and O'Hanesian, M.~A. (1995).
\newblock An analysis for menstrual data with time-varying covariates.
\newblock {\em Statistics in Medicine} {\bf 14,} 1843--1857.

\bibitem[\protect\citeauthoryear{Royston and Abrams}{Royston and
  Abrams}{1980}]{Royston1980}
Royston, J.~P. and Abrams, R.~M. (1980).
\newblock An objective method for detecting the shift in basal body temperature
  in women.
\newblock {\em Biometrics} {\bf 36,} 217--224.

\bibitem[\protect\citeauthoryear{Ruckdeschel and Kohl}{Ruckdeschel and
  Kohl}{2014}]{Ruckdeschel2014}
Ruckdeschel, P. and Kohl, M. (2014).
\newblock {General purpose convolution algorithm in S4 classes by means of
  FFT}.
\newblock {\em Journal of Statistical Software} {\bf 59,} 1--25.

\bibitem[\protect\citeauthoryear{S\"{a}rkk\"{a}}{S\"{a}rkk\"{a}}{2013}]{Sarkka2013}
S\"{a}rkk\"{a}, S. (2013).
\newblock {\em {Bayesian Filtering and Smoothing}}.
\newblock Cambridge University Press, Cambridge.

\bibitem[\protect\citeauthoryear{Scarpa}{Scarpa}{2014}]{Scarpa2010}
Scarpa, B. (2014).
\newblock Probabilistic and statistical models for conception.
\newblock {\em Wiley StatsRef: Statistics Reference Online}.

\bibitem[\protect\citeauthoryear{Scarpa and Dunson}{Scarpa and
  Dunson}{2009}]{Scarpa2009}
Scarpa, B. and Dunson, D.~B. (2009).
\newblock Bayesian hierarchical functional data analysis via contaminated
  informative priors.
\newblock {\em Biometrics} {\bf 65,} 772--780.

\bibitem[\protect\citeauthoryear{Schwartz, Macdonald, and Heuchel}{Schwartz
  et~al.}{1980}]{Schwartz1980}
Schwartz, D., Macdonald, P. D.~M., and Heuchel, V. (1980).
\newblock Fecundability, coital frequency and the viability of ova.
\newblock {\em Population Studies} {\bf 34,} 397--400.

\bibitem[\protect\citeauthoryear{Sundaram, Buck~Louis, and Kim}{Sundaram
  et~al.}{2014}]{Sundaram2010}
Sundaram, R., Buck~Louis, G.~M., and Kim, S. (2014).
\newblock Statistical modeling of human fecundity.
\newblock {\em Wiley StatsRef: Statistics Reference Online}.

\bibitem[\protect\citeauthoryear{Weinberg and Dunson}{Weinberg and
  Dunson}{2000}]{Weinberg2000}
Weinberg, C.~R. and Dunson, D.~B. (2000).
\newblock Some issues in assessing human fertility.
\newblock {\em Journal of the American Statistical Association} {\bf 95,}
  300--303.

\bibitem[\protect\citeauthoryear{Zhou}{Zhou}{2006}]{Zhou2006}
Zhou, H. (2006).
\newblock Statistical models for human fecundability.
\newblock {\em Statistical Methods in Medical Research} {\bf 15,} 181--194.

\end{thebibliography}

\begin {table*}[p]
  \centering
  \caption{Summary of the menstrual cycle data. Note that users who provided records over several years can be counted in multiple age groups.}
\scalebox{0.7}[0.7]{
\centering
	\begin{tabular}{lcccccccc}
	\toprule
	 	   & \multicolumn{8}{c}{Age (years)} \\
		   & 15--19 & 20--24 & 25--29 & 30--34 & 35--39 & 40--44 & 45--49 & 50--54\\	 
	\midrule
  Entire data set\\
		\hspace{2em} No. of subjects & 118 & 542 & 1,020 & 1,090 & 781 & 364 & 134 & 19\\
    \hspace{2em} No. of cycles	& 411 & 2,636 & 5,087 & 	6,496 & 5,903 & 3,479 & 1,438 & 172 \\
		\hspace{2em} Range of cycle length	& (15, 53) & (21, 50) & (24, 48) & (24, 44) & (23, 40) & (21, 41) & 	(20, 56) & 	(16, 87) \\
		\hspace{2em} Mean of cycle length & 30.4 & 31.8 & 31.5 & 30.4 & 29.2 & 27.9 & 28.8 & 32.3 \\
		\hspace{2em} Median of cycle length	 & 30 & 31 & 31 & 30 & 29 & 27 & 27 & 28 \\
		\hspace{2em} SD of cycle length & 6.7 & 5.6 &	4.7 & 4.1 & 3.4 & 3.6 & 5.8 & 12.7 \\
    \hspace{2em} No. of observations & 12,660 & 84,538 & 161,531 & 198,871 & 173,563 & 97,610 & 41,647 &	5,582 \\
		\hspace{2em} Percentage of missing observations & 13.7 & 15.8 & 14.0 & 10.8 & 10.5 & 9.0 & 8.8 & 4.0 \\
	
	 	Data for parameter estimation\\
		 \hspace{2em} No. of subjects & 97 & 300 & 300 & 300 & 300 & 148 & 84 &17\\
		 \hspace{2em} No. of cycles & 300 & 300 & 300 & 300 & 300 & 300 & 300 & 120\\
		 \hspace{2em} Range of cycle length	 & (15, 53) & (21, 49) & (24, 47) & (24, 43) & (24, 40) & (22, 40) & (20, 51) & (16, 87)\\
		 \hspace{2em} Mean of cycle length	& 30.4 & 31.7 & 32.0 & 30.8 & 29.8 & 28.1 & 28.1 & 31.4\\
		 \hspace{2em} Median of cycle length & 30 & 31 & 31 & 30 & 29 & 27 &27 & 28\\
		 \hspace{2em} SD of cycle length & 6.7 & 5.4 & 4.8 & 4.4 & 3.8 & 3.6 & 4.8 & 11.8\\
     \hspace{2em} No. of observations & 9,281 & 9,822 & 9,887 & 9,537 & 9,247 & 8,701 & 8,674 & 3,819\\
		 \hspace{2em} Percentage of missing observations & 13.5 & 16.9 & 16.0 & 14.1 & 12.8 & 	9.0 & 9.1 & 3.7 \\

		 Data for predictive accuracy estimation\\
		 \hspace{2em} No. of subjects & 64 & 150 & 150 & 150 & 150 &101 & 70 & 15\\
		 \hspace{2em} No. of cycles & 111 & 150 & 150 & 150 & 150 & 150 & 150 & 52\\
		 \hspace{2em} Range of cycle length	 & (16, 49) & (21, 49) & (24, 45) & (24, 42) & (24, 40)  & (22, 40) & (21, 56) & (17, 77)\\
		 \hspace{2em} Mean of cycle length & 30.5 & 33.4 & 32.1 & 31.0 & 29.7 & 27.9 & 29.6 & 34.3\\
		 \hspace{2em} Median of cycle length & 30 & 32 & 31 & 31 & 29 & 27 & 27 & 29.5\\
		 \hspace{2em} SD of cycle length &6.7 & 6.4 & 4.6 & 3.9 & 3.8 & 3.9 & 6.7 & 14.6\\
     \hspace{2em} No. of observations & 3,481 & 5,162 & 4,962 & 4,803 & 4,610 & 4,333 & 4,577 &1,827\\
		 \hspace{2em} Percentage of missing observations & 13.9 & 17.4 & 16.3 & 14.5 & 12.7 & 	10.1 & 8.6 & 4.4 \\
	\bottomrule
	\end{tabular}
	
	}
	\label{table:data}
\end{table*}

\begin{sidewaystable*}[p]
  \centering
  \caption{Maximum likelihood estimates of the parameters and their 95\% confidence intervals (in brackets).
$\dagger$ Confidence intervals were not calculated because the Hessian of log-likelihood was singular.}
\scalebox{0.7}[0.7]{
\centering
	\begin{tabular}{lcccccccc}
	\toprule
  & \multicolumn{8}{c}{Age (years)} \\
		   & 15--19 & 20--24 & 25--29 & 30--34 & 35--39 & 40--44 & 45--49 & 50--54\\	 
	\midrule
				System model\\
		\hspace{2em} First stage \\
		\hspace{4em} $\alpha_1$	&0.632 &0.942 &0.871 &1.316 &0.952 &1.000 & 0.644&0.054\\    
    \hspace{4em} & (0.475, 0.842) & (0.691, 1.283) & (0.672, 1.128) & (0.665, 2.602) & (0.755, 1.201) & $\dagger$ & (0.511, 0.811) & (0.036, 0.080) \\
		\hspace{4em} $\beta_1$	&34.923 &52.320 &43.145 &64.430 &40.455 & 42.920& 26.902 &1.853\\ 
    \hspace{4em} & (25.792, 47.287) & (37.764, 72.488) & (32.794, 56.762) & (33.701, 123.177) & (31.706, 51.618) & $\dagger$ & (21.017, 34.434) & (0.970, 3.542) \\
		\hspace{2em} Second stage \\
		\hspace{4em} $\alpha_2$ &0.216 &0.271 &0.350 &0.364 &0.533 & 0.398& 0.334&0.177\\    
    \hspace{4em} & (0.181, 0.258) & (0.228, 0.321) & (0.293, 0.418) & (0.304, 0.435) & (0.413, 0.688) & (0.337, 0.469) & (0.279, 0.399) & (0.126, 0.249)\\
		\hspace{4em} $\beta_2$	&1.837 &2.865 &5.141 &5.218 &8.669 & 5.783& 4.472&2.170\\
    \hspace{4em}  & (1.330, 2.536) & (2.183, 3.759) & (4.013, 6.586) & (4.020, 6.772) & (6.245, 12.033) & (4.688, 7.134) & (3.455, 5.789) & (1.186, 3.969)\\
		Observation model\\			
		\hspace{2em} First stage \\
		\hspace{4em} $\mu_1$ &$-0.040$ &$-0.040$ &$-0.029$ & $-0.012$ &$-0.023$ &$-0.024$ &$-0.018$ &$-0.054$\\ 
    \hspace{4em} &$(-0.047, -0.034)$ & $(-0.047, -0.033)$ & $(-0.035, -0.022)$ & $(-0.018, -0.005)$ & $(-0.031, -0.015)$ & $(-0.029, -0.019)$ & $(-0.024, -0.012)$ & $(-0.063, -0.044)$\\
    \hspace{4em} $\sigma_1$	&0.231 &0.239 &0.228 &0.217 &0.252 & 0.207&0.203&0.226\\    
  \hspace{4em}  & (0.226, 0.235) & (0.235, 0.244) & (0.223, 0.233) & (0.213, 0.221) & (0.247, 0.257) & (0.204, 0.211) & (0.199, 0.207) & (0.220, 0.233) \\
		
		\hspace{2em} Second stage\\
		\hspace{4em} $\mu_2$	&0.371 &0.374 &0.369 &0.377 &0.363 & 0.333& 0.325&0.345\\    
    \hspace{4em} &(0.359, 0.382) & (0.363, 0.384) & (0.360, 0.378) & (0.368, 0.387) & (0.354, 0.372) & (0.325, 0.341) & (0.317, 0.333) & (0.330, 0.359)\\
		\hspace{4em} $\sigma_2$	&0.247 &0.224 &0.220 &0.223 &0.205 & 0.199& 0.196&0.216\\    
    \hspace{4em} &(0.239, 0.255) & (0.217, 0.231) &(0.213, 0.227) &(0.217, 0.230) &(0.198, 0.211) &(0.193, 0.204) &(0.191, 0.202) &(0.207, 0.225)\\
    Shift in BBT \\
		\hspace{4em} $\mu_2 - \mu_1$	&0.411 &0.413 &0.398 &0.389 &0.386 & 0.357& 0.342&0.398\\ 
    &(0.399, 0.423) &(0.402, 0.425) &(0.387, 0.408) &(0.378, 0.400) &(0.375, 0.397) &(0.348, 0.367) &(0.333, 0.352) &(0.382, 0.414)\\
	\bottomrule
	\end{tabular}
	}
	\label{table:param}
\end{sidewaystable*}

\begin{sidewaystable*}[p]
    \centering
  \caption{A list of models for which accuracy of the prediction of menstruation onset was examined. Note that in our application, the index $i$ denotes cycles rather than subjects.}
  \scalebox{0.7}[0.7]{
    \begin{tabular}{lllll}
      \toprule
      Model & Code & System model & Observation model & Sequential Bayesian prediction  \\
      \midrule
      Fully explicit & FE & $\theta_{it} = \theta_{i,t-1} + \epsilon_{it}; \epsilon_{it} \sim \textrm{Gamma}\left\{\alpha\left(\theta_{i,t-1}\right), \beta\left(\theta_{i,t-1}\right) \right\}$ & $y_{it} = \mu\left(\theta_{it}\right) + e_{it}; e_{it} \sim \textrm{Normal}\left\{0, \sigma^2\left(\theta_{it}\right)\right\}$ & yes  \\
      && $\left\{\alpha\left(\theta_{i,t-1}\right), \beta\left(\theta_{i,t-1}\right) \right\} = \begin{cases} 
      (\alpha_{1}, \beta_{1}) \hspace{1em} \textrm{when} \hspace{1em}
      \theta_{i,t-1} \in \Theta_1 \\
      (\alpha_{2}, \beta_{2})  \hspace{1em} \textrm{when} \hspace{1em}
      \theta_{i,t-1} \in \Theta_2 \\
    \end{cases}$ & $\left\{\mu\left(\theta_{it}\right), \sigma^2\left(\theta_{it}\right)\right\} = \begin{cases} 
      (\mu_{1}, \sigma^2_{1}) \hspace{1em} \textrm{when} \hspace{1em}
      \theta_{it} \in \Theta_1 \\
      (\mu_{2}, \sigma^2_{2})  \hspace{1em} \textrm{when} \hspace{1em}
      \theta_{it} \in \Theta_2 \\
    \end{cases}$ & \\
    Restricted explicit & RE & $\theta_{it} = \theta_{i,t-1} + \epsilon_{it}; \epsilon_{it} \sim \textrm{Gamma}(\alpha, \beta)$ & $y_{it} = \mu\left(\theta_{it}\right) + e_{it}; e_{it} \sim \textrm{Normal}\left\{0, \sigma^2\left(\theta_{it}\right)\right\}$ & yes  \\
      &&& $\left\{\mu\left(\theta_{it}\right), \sigma^2\left(\theta_{it}\right)\right\} = \begin{cases} 
      (\mu_{1}, \sigma^2_{1}) \hspace{1em} \textrm{when} \hspace{1em}
      \theta_{it} \in \Theta_1 \\
      (\mu_{2}, \sigma^2_{2})  \hspace{1em} \textrm{when} \hspace{1em}
      \theta_{it} \in \Theta_2 \\
    \end{cases}$ & \\
      Implicit 1 & I1 & $\theta_{it} = \theta_{i,t-1} + \epsilon_{it}; \epsilon_{it} \sim \textrm{Gamma}(\alpha, \beta)$ & $y_{it} = \mu\left(\theta_{it}\right) + e_{it}; e_{it} \sim \textrm{Normal}\left(0, \sigma^2\right)$ & yes  \\
      &&& $\mu\left(\theta_{it}\right) = a + b_1\cos2\pi\theta_{it} + c_1\sin2\pi\theta_{it}$ & \\
      Implicit 2 & I2 & $\theta_{it} = \theta_{i,t-1} + \epsilon_{it}; \epsilon_{it} \sim \textrm{Gamma}(\alpha, \beta)$ & $y_{it} = \mu\left(\theta_{it}\right) + e_{it}; e_{it} \sim \textrm{Normal}\left(0, \sigma^2\right)$ & yes  \\
      &&& $\mu\left(\theta_{it}\right) = a + \sum_{m=1}^2 b_m\cos2m\pi\theta_{it} + c_m\sin2m\pi\theta_{it}$ & \\
      Implicit 3 & I3 & $\theta_{it} = \theta_{i,t-1} + \epsilon_{it}; \epsilon_{it} \sim \textrm{Gamma}(\alpha, \beta)$ & $y_{it} = \mu\left(\theta_{it}\right) + e_{it}; e_{it} \sim \textrm{Normal}\left(0, \sigma^2\right)$ & yes  \\
      &&& $\mu\left(\theta_{it}\right) = a + \sum_{m=1}^3 b_m\cos2m\pi\theta_{it} + c_m\sin2m\pi\theta_{it}$ & \\
      Calendar calculation & C & NA & NA & no  \\
      \bottomrule
    \end{tabular}
  }
  \label{table:models}
\end{sidewaystable*}

\begin {table*}[p]
\centering
\caption{Summary of lengths of two stages.
  Monophasic cycles were defined as cycles in which the length of the second stage was
  estimated to be less than three days, based on the phase identification method described in
  section \ref{sect:phase}.
  The smoothed probability distribution of the menstrual cycle phase was used to
  determine the stage of the menstrual cycle.
}
\scalebox{0.7}[0.7]{
\centering
	\begin{tabular}{lcccccccc}
	\toprule
  & \multicolumn{8}{c}{Age (years)} \\
		   & 15--19 & 20--24 & 25--29 & 30--34 & 35--39 & 40--44 & 45--49 & 50--54\\	 
	\midrule
		All cycles\\
		\hspace{2em} First stage \\
		\hspace{4em} Mean	& 21.4 & 22.0 & 20.6 & 19.5 & 18.5 & 17.4 & 18.3 & 23.2 \\
		\hspace{4em} Median	& 21 & 21 & 20 & 19 & 18 & 17 & 17 & 18 \\
		\hspace{4em} SD	& 7.6 & 6.6 & 5.7 & 5.0 & 4.4 & 4.9 & 7.2 & 15.2 \\
		\hspace{2em} Second stage \\
		\hspace{4em} Mean	& 9.1 & 9.8 & 10.9 & 10.9 & 10.7 & 10.5 & 10.4 & 9.1 \\
		\hspace{4em} Median	& 10 & 10 & 11 & 11 & 11 & 11 & 11 & 10 \\
		\hspace{4em} SD	& 5.3 & 4.4 & 3.9 & 3.6 & 3.4 & 3.8 & 4.5 & 6.5 \\
				
		Without monophasic cycles \\							

		\hspace{2em} First stage \\
		\hspace{4em} Mean	& 20.4 & 21.3 & 20.2 & 19.2 & 18.2 & 16.9 & 17.4 & 18.3 \\
		\hspace{4em} Median	& 20 & 21 & 20 & 19 & 18 & 16 & 16 & 16 \\
		\hspace{4em} SD	& 7.1 & 6.2 & 5.4 & 4.7 & 4.1 & 4.4 & 6.3 & 10.0 \\
		\hspace{2em} Second stage \\
		\hspace{4em} Mean	& 10.5 & 10.6 & 11.3 & 11.1 & 11.0 & 11.0 & 11.2 & 11.9 \\
    \hspace{4em} Median	& 10 & 11 & 11 & 11 & 11 & 11 & 11 & 11.5 \\
		\hspace{4em} SD	& 4.4 & 3.7 & 3.4 & 3.3 & 3.0 & 3.2 & 3.8 & 4.9 \\
		Percentage of monophasic cycles 
		& 15.6 & 8.5 & 3.8 & 2.8 & 2.9 & 4.9 & 8.1 & 24.4 \\
	\bottomrule
	\end{tabular}
	}
  \label{table:stage_summary}
\end{table*}

\begin{table*}[p]
  \centering
  \caption{Mathematical notations used in Appendix \ref{sect:appa}.}
  \scalebox{0.7}[0.7]{
    \hspace{-4em}
    \begin{tabular}{lp{10cm}} \toprule
    Definition & Description \\
    \midrule
    $\epsilon^{(m)}_{it}$ & 
    The advance of the menstrual phase between $t-1$ and $t$, given that subject $i$ was in the
    first ($m=1$) or second ($m=2$) stage of the cycle at $t-1$.  \\
    $G(\hspace{.1em}\cdot\hspace{.1em} ; s, r)$ &
    Distribution function of the gamma distribution with shape parameter $s$
    and rate parameter $r$.  \\
    $g(\hspace{.1em}\cdot\hspace{.1em} ; s, r)$ &
    Probability density function of the gamma distribution with shape parameter $s$
    and rate parameter $r$. \\
    $\left\{f_1(\xi_1) * f_2(\xi_2)\right\}(x) = \int f_1(x - t; \xi_1)f_2(t; \xi_2)dt$ & 
      Convolution of probability density functions $f_1(\cdot; \xi_1)$ and $f_2(\cdot; \xi_2)$,
      which respectively has a vector of parameters $\xi_1$ and $\xi_2$. \\
      $u_{ij}= \sum_{r=1}^{j-1} \epsilon^{(1)}_{i,t+r}
      \hspace{.5em} \textrm{for} \hspace{.5em} j > 1$ & 
      The advance of the phase accumulated up to the final day of the first stage ($t+j-1$). \\
      $\pi_{u_{ij}}(x)= g\left\{ x; (j-1)\alpha_{i1}, \beta_{i1} \right\}$ & 
      Probability density function of the distribution which $u_{ij}$ follows. \\
      $\pi^{*}_{u_{ij}}(x; a, b) = \begin{cases}
        \frac{\pi_{u_{ij}}(x)}{\int_{a}^{b} \pi_{u_{ij}}(y)dy}
        \hspace{.5em} \textrm{for} \hspace{.5em} (a < x < b) \\
      0 \hspace{.5em} \textrm{otherwise} \end{cases}$ & 
      Probability density function of the distribution which $u_{ij}$,
      given $a < u_{ij} < b$, follows. \\
      $v_{ij} = \begin{cases}
        \epsilon^{(1)}_{i,t+1}
        \hspace{.5em} \textrm{for} \hspace{.5em} j=1 \\
        u_{ij} + \epsilon^{(1)}_{i,t+j}
      \hspace{.5em} \textrm{for} \hspace{.5em} j>1,
      \textrm{given} \hspace{.5em} 0 < u_{ij} < \thlfh \end{cases}$ & 
      The advance of the phase accumulated up to the day 
      at which subject $i$ leaves the first stage ($t+j$). \\
      $\pi_{v_{ij}}(x) = \begin{cases}
        g\left(x; \alpha_{i1}, \beta_{i1} \right)
        \hspace{.5em} \textrm{for} \hspace{.5em} j=1 \\
        \left[ \pi^*_{u_{ij}}\left\{ 0, \thlfh \right\} * g(\alpha_{i1}, \beta_{i1}) \right](x)
      \hspace{.5em} \textrm{for} \hspace{.5em} j>1 \end{cases}$ & 
      Probability density function of the distribution which $v_{ij}$ follows. \\
      $\pi^{*}_{v_{ij}}(x; a, b) = \begin{cases}
        \frac{\pi_{v_{ij}}(x)}{\int_{a}^{b} \pi_{v_{ij}}(y)dy}
        \hspace{.5em} \textrm{for} \hspace{.5em} (a < x < b) \\
      0 \hspace{.5em} \textrm{otherwise} \end{cases}$ & 
      Probability density function of the distribution which $v_{ij}$,
      given $a < v_{ij} < b$, follows. \\
      $w_{ijk} = v_{ij} + \sum_{r=j+1}^{k-1}\epsilon^{(2)}_{i,t+r}
      \hspace{.5em} \textrm{for} \hspace{.5em} j \geq 1, k > j + 1,
      \textrm{given} \hspace{.5em} \thlfh < v_{ij} < \thllh$ &
      The advance of the phase accumulated up to the day immediately before the menstruation onset
      ($t+k-1$). \\
      $\pi_{w_{ijk}}(x) = \left[ \pi^*_{v_{ij}}\left\{ \thlfh, \thllh \right\} * 
        g\left\{ (k-j-1)\alpha_{i2}, \beta_{i2} \right\} \right](x)$ & 
      Probability density function of the distribution which $w_{ijk}$ follows.   \\
      $\pi^{*}_{w_{ijk}}(x; a, b) = \begin{cases}
        \frac{\pi_{w_{ijk}}(x)}{\int_{a}^{b} \pi_{w_{ijk}}(y)dy}
        \hspace{.5em} \textrm{for} \hspace{.5em} (a < x < b) \\
      0 \hspace{.5em} \textrm{otherwise} \end{cases}$ & 
      Probability density function of the distribution which $w_{ijk}$,
      given $a < w_{ijk} < b$, follows. \\
    \bottomrule
  \end{tabular}
  }
  \label{table:app}
\end{table*}

\end{spacing}

\end{document}